\documentclass[12pt,a4paper]{article}

\usepackage[left=5em]{geometry}
\usepackage{amsmath}

\title{Trapping Horizons in the Sultana-Dyer Space-Time}
\author{\textbf{Cheng-Yi Sun\footnote{cysun@mails.gucas.ac.cn; ddscy@163.com}\ $^{,a}$\
}\\ \\
 {$^a$\small Institute of Modern Physics, Northwest University,}\\
     \small Xian 710069, P.R. China.}

\begin{document}
\maketitle
\begin{abstract}
The Sultana-Dyer space-time is suggested as a model describing a
black hole embedded in an expanding universe. Recently, in
\cite{0705.4012}, its global structure is analyzed and the trapping
horizons are shown. In the paper, by directly calculating the
expansions of the radial null vector fields normal to the space-like
two-spheres foliating the trapping horizons, we find that the
trapping horizon outside the event horizon in the Sultana-Dyer
space-time is a past trapping horizon. Further, we find that the
past trapping horizon is an outer, instantaneously degenerate or
inner trapping horizon accordingly when the radial coordinate is
less than, equal to or greater than some value.

\end{abstract}



\section{Introduction}
Isolated black holes have been investigated in great depth and
detail for more than forty years. On the other hand, black holes
embedded in the background of an expanding universe are also
important and even more relativistic. Some models describing black
holes embedded in expanding universes have been suggested. In
\cite{McVittie}, the McVittie model is suggested. In \cite{Swiss
cheese}, the model of Swiss cheese black holes  is shown. In
\cite{Vaidya}, the author suggested the Vaidya's spacetime
describing a FRW universe with a Schwarzschild-like black hole that
does not expand with the rest of the universe. In \cite{Thakurta},
the Thakurta's black hole is shown. In \cite{Sultana and Dyer}, the
Sultana-Dyer black hole model is suggested. Recently, in
\cite{McClure and Dyer,0707.1350}, the authors suggested new
solutions describing black holes embedded in the expanding universe.

It is interesting to analyze the structures of these cosmological
black hole models. The McVittie space-time has been analyzed
extensively by Nolan \cite{Nolan}. But for other models, further
analysis is still needed. In \cite{0705.4012}, the global structure
of the Sultana-Dyer model is discussed, and its two trapping
horizons are given by calculating the product of the expansions of
the outgoing and ingoing future null vector fields normal to the
space-like 2-spheres foliating the trapping horizons. But in
\cite{0705.4012}, the expansions themselves are not given. In the
paper, we will calculate directly the expansions, and then discuss
some properties of the trapping horizons.

This paper is organized as follows. In Section \ref{Sultana-Dyer},
we recall the Sultana-Dyer model. In Section \ref{trapped surfaces},
the expansions of the outgoing and ingoing null normal vector fields
is calculated. Section \ref{Conclusion} contains the summary.

\section{the Sultana-Dyer Model}
\label{Sultana-Dyer}

The Sultana-Dyer space-time is described by the metric \cite{Sultana
and Dyer}
\begin{equation}
  \label{SDMetric}
  ds^2=a^2(\eta)\left[-(1-\frac{2m}{r})d\eta^2+\frac{4m}{r}d\eta
  dr+(1+\frac{2m}{r})dr^2+r^2d\Omega^2\right],
\end{equation}
where $m$ is a constant, $d\Omega^2=d\theta^2+\sin^2\theta
d\varphi^2$, $a(\eta)=(\eta/\eta_*)^2$ and $\eta_*$ is a constant.
The metric is sourced by the mixture of a massive dust and a null
dust with the energy-momentum \cite{Sultana and Dyer}
\begin{equation}
  \label{SDEMTensor}
  T^{ab}=T_{(1)}^{ab}+T_{(2)}^{ab},
\end{equation}
where $T_{(1)}^{ab}=\rho_1u^au^b$ describes an ordinary massive dust
with density $\rho_1$, and $T_{(2)}^{ab}=\rho_2k^ak^b$ describes a
null dust with $k^ak_a=0$ and density $\rho_2$
\cite{0705.4012,0907.4473}.

We define a new time coordinate $t$ as $dt=ad\eta$. Then the metric
(\ref{SDMetric}) is rewritten as
\begin{equation}
  \label{SDMetriT}
  ds^2=-(1-\frac{2m}{r})dt^2+\frac{4m}{r}a(t)dt
  dr+(1+\frac{2m}{r})a^2(t)dr^2+a^2(t)r^2d\Omega^2,
\end{equation}
where $a(t)=(t/t_*)^{2/3}$ and $t_*=\eta_*/3$.

\section{the Trapping Horizons}
\label{trapped surfaces}

The radial null geodesics for the metric (\ref{SDMetriT}) can be
easily found by setting $ds=d\Omega=0$. We have
\begin{equation}
  \label{drdt}
  \frac{dr}{dt}=\frac{\pm1-\frac{2m}{r}}{a(1+\frac{2m}{r})},
\end{equation}
where plus sign denotes the outgoing geodesics and minus sign
denotes the ingoing geodesics. Thus we can find the outgoing and
ingoing future-pointing radial null vector fields $l^a$ and $n^a$
with components
\begin{eqnarray}
  \label{l}
  l^a&=&\alpha(t,r)\left(1,\frac{1-\frac{2m}{r}}{a(1+\frac{2m}{r})},0 ,0\right),\\
  \label{n}
  n^a&=&\beta(t,r)\left(1,-\frac{1}{a},0 ,0\right),
\end{eqnarray}
where the positive functions $\alpha$ and $\beta$ are chosen to
ensure $l^a\nabla_a l^a=n^a\nabla_a n^a=0$ and $l^an_a=-1$. Then for
a space-like 2-sphere $S^2$ which may be locally defined by
$t=const.$ and $r=const.$, the expansions $\theta_l$ and $\theta_n$
associated with the vector fields $l^a$ and $n^a$ respectively can
be defined as \cite{0809.3850}
\begin{eqnarray}
  \label{DefThetal}
  \theta_l&=&\nabla_al^a+(n^al^b+l^an^b)\nabla_al_b,\\
  \label{DefThetan}
  \theta_n&=&\nabla_an^a+(n^al^b+l^an^b)\nabla_an_b.
\end{eqnarray}
As the result of the spherical symmetry \cite{0810.2712,0908.3101},
we have
\begin{eqnarray}
  \label{thetal}
  \theta_l&=&\frac{2}{R}l(R)=\frac{2\alpha}{R}\left(HR+\frac{1-\frac{2m}{r}}{1+\frac{2m}{r}}\right),\\
  \label{thetan}
  \theta_n&=&\frac{2}{R}n(R)=\frac{2\beta}{R}(HR-1),
\end{eqnarray}
where $R\equiv ar$ and $H\equiv\frac{da/dt}{a}$.

Obviously, outside the event horizon $r=2m$, there exists a trapping
horizon described by
\begin{equation}
  \label{H1}
  R=H^{-1},
\end{equation}
which is just the trapping horizon $r_2=\eta/2$ given in
\cite{0705.4012}. We use $H_1$ to denote this trapping horizon. And
inside the event horizon there is the other trapping horizon
described by
\begin{equation}
  \label{H2}
  r=\frac{-(1+2m\dot{a})+\sqrt{(1+2m\dot{a})^2+8m\dot{a}}}{2\dot{a}},
\end{equation}
which is just the trapping horizon
$r_1=-m+(-\eta+\sqrt{\eta^2+24m\eta+16m^2})/4$ given in
\cite{0705.4012}. Here $\dot{a}\equiv da/dt$. We use $H_2$ to denote
this trapping horizon.

Then the trapping horizon $H_1$ is a past trapping horizon, since
at $H_1$
\[
  \theta_n=0, \quad \theta_l>0.
\]
And the trapping horizon $H_2$ is a future trapping horizon, since
at $H_2$
\[
  \theta_l=0, \quad \theta_n<0.
\]

Further, we find at the trapping horizon $H_1$
\begin{equation}
  \label{dThetandl}
  l^a\nabla_a\theta_n|_{H_1}=\frac{4\alpha\beta}{9t^2}\frac{1-\frac{6m}{r}}{1+\frac{2m}{r}}.
\end{equation}
Then $l^a\nabla_a\theta_n|_{H_1}$ is negative, zero or positive
accordingly when $r$ is less than, equal to or greater than $6m$.
Thus the trapping horizon $H_1$ is an outer, instantaneously
degenerate or inner trapping horizon accordingly when $r$ is less
than, equal to or greater than $6m$.

Similarly, we find at the trapping horizon $H_2$
\begin{equation}
  \label{dThetaldn}
   n^a\nabla_a\theta_l|_{H_2}=-2\alpha\beta\Big\{\frac{2}{9t^2}+\frac{H}{R}+\frac{1}{R^2}\frac{4m/r}{(1+2m/r)^2}\Big\}\\
\end{equation}
Then $n^a\nabla_a\theta_l|_{H_2}$ is always negative. Thus the
trapping horizon $H_2$ is an outer trapping horizon.

\section{Summary}
\label{Conclusion}

The issue on the solutions of the Einstein equation describing black
holes embedded in the FRW universe is very important. Recently, a
new solution, the Sultana-Dyer model, is suggested \cite{Sultana and
Dyer}. In \cite{0705.4012}, the global structure the Sultana-Dyer
space-time is discussed and two trapping horizons are given. In the
paper, by calculating the expansions associated with the outgoing
and ingoing future radial null vector fields, we firstly check the
existence of the two trapping horizons shown in \cite{0705.4012}.
Then we find that the trapping horizon $H_1$ outside the event
horizon is a past trapping horizon, and the trapping horizon $H_2$
inside the event horizon is a future trapping horizon. Further we
calculate the derivative of the expansion associated with the
ingoing null vector along the outgoing null vector at the trapping
horizon $H_1$. Then we find the trapping horizon $H_1$ is an outer,
instantaneously degenerate or inner trapping horizon accordingly
when $r$ is less than, equal to or greater than $6m$. Similarly, we
find the trapping horizon $H_2$ is an outer trapping horizon.

Here we note that in \cite{0907.4473} the different conclusions are
given, in which the authors claimed that there are two trapping
horizon outside the event horizon in the Sultana-Dyer space-time. In
fact, as having been noted in \cite{0908.3101}, the metric on which
the conclusions in \cite{0907.4473} are based does not describe  the
Sultana-Dyer space-time. Thus it is natural that the different
conclusions are obtained in \cite{0907.4473}. Then the singular
surface at $\tilde{r}=2m$ (See \cite{0907.4473} for definition)
shown in \cite{0907.4473} does not belong to the Sultana-Dyer, too.
The scalar curvature $R_c$ of the Sultana-Dyer model can be easily
obtained
\begin{equation}
  \label{Rc}
  R_c=\frac{6[r(2m+r)a''-2ma']}{a^3r^2},
\end{equation}
where $a'\equiv\frac{da}{d\eta}$, $a''\equiv\frac{d^2a}{d\eta^2}$.
Then we know that in the Sultana-Dyer model, there is no singularity
except at $r=0$.

\section*{Acknowledgments}
This work is supported by the Natural Science Foundation of
Northwest University of China under Grant No. NS0927.

\end{document}